\documentclass[showpacs,aps,graphicx]{revtex4}%,,twocolumn,preprint
\usepackage{amssymb}
\usepackage{graphicx}

\begin{document}

\title{Coulomb effects on the formation of proton halo nuclei}
\author{Yu-Jie Liang$^{1,2,3}$, Yan-Song Li$^{4}$, Fu-Guo Deng$^{1,2,3}$,
Xi-Han Li$^{1,2,3}$, Bao-An Bian$^{1,2,3}$,  Feng-Shou
Zhang$^{1,2,3}$, Zu-Hua Liu$^{1,2,3,5}$, and Hong-Yu
Zhou$^{1,2,3}$\footnote{ Email: zhy@bnu.edu.cn}}.
\address{$^1$ The Key
Laboratory of Beam Technology and Material Modification of Ministry
of Education, Beijing Normal University, Beijing 100875,
People's Republic of China\\
$^2$ Institute of Low Energy Nuclear Physics, and Department of
Material Science and Engineering, Beijing Normal University, Beijing
100875,
People's Republic of China\\
$^3$ Beijing Radiation Center, Beijing 100875,  People's Republic of
China\\
$^4$ Department of Physics, Tsinghua University, Beijing 100084,
People's Republic of China\\
$^5$ China Institute of Atomic Energy, Beijing 102413, People's
Republic of China}

\begin{abstract}
The exotic structures in the 2$s_{1/2}$ states of five pairs of
mirror nuclei $^{17}$O-$^{17}$F, $^{26}$Na-$^{26}$P,
$^{27}$Mg-$^{27}$P, $^{28}$Al-$^{28}$P and $^{29}$Si-$^{29}$P are
investigated with the relativistic mean-field (RMF) theory and the
single-particle model (SPM) to explore the role of the Coulomb
effects on the proton halo formation. The present RMF calculations
show that the exotic structure of the valence proton is more obvious
than that of the valence neutron of its mirror nucleus, the
difference of exotic size between each mirror nuclei becomes smaller
with the increase of mass number A of the mirror  nuclei and the
ratios of the valence proton and valence neutron  root-mean-square
(RMS) radius to the matter radius in each pair of mirror nuclei all
decrease linearly with the increase of A. In order to interpret
these results, we analyze two opposite effects of Coulomb
interaction on the exotic structure formation with SPM and find that
the contribution of the energy level shift is more important than
that of the Coulomb barrier for light nuclei. However, the hindrance
of the Coulomb barrier becomes more obvious with the increase of A.
When A is larger than 34, Coulomb effects on the exotic structure
formation will almost become zero because its two effects counteract
with each other.
\end{abstract}

\date{\today}
\maketitle

\section{Introduction}

Nuclear halo is a kind of exotic structures in which the nuclear
matter distribution extends to large radii because of its weakly
bound character. Halo nuclei have been extensively investigated both
experimentally and theoretically for decades
\cite{Tani85,Mitt87,Hans87,
Suzu88,Sain89,Liat90,Joha90,Bert91,Haye91,Zhuk93,Thom93,Ren96,Morl97,Ren99,Liu01,
Guim95,Axel96,Aoya00, Auer00,Zhang03,Chen05,Liang06}. However, up to
date most of the halo nuclei confirmed are the neutron halo. The
proton halos observed are rather scarce. Since $^{8}$B and $^{17}$Ne
are predicted as proton halos \cite{Riis93,Al-kh96,Esbe96,Cole98},
the first excited state of $^{17}$F\cite{Morl97,Ren98} and
proton-rich isotopes P and S \cite{Ren96,Brow96,Chen98,Navi98,Ren99}
are predicted as the proton halos and some of them are probed
experimentally.

There is a popular opinion that it is more difficult to form proton
halo because the Coulomb barrier hinders the proton penetrating into
the out region of nucleus. However, this may be not true actually in
some cases. As will be seen below, for lighter nuclei the proton
halo is easier to occur as compared to neutron halo due to the
Coulomb interaction. The Coulomb interaction has two effects on the
formation of nuclear halo. One of these effects is that it makes the
energy level shift closer to the Fermi level thus facilitating the
penetration of the valence proton beyond the range of nuclear force.
On the other hand, the Coulomb barrier of the proton hinders the
formation of halo structure. In the case of lighter nuclei, the
former effect is more important than the later one.

Because nuclear force is nearly charge-independent, the structure
difference between mirror nuclei should mainly come form the effects
of the Coulomb interaction. Therefore, the mirror nuclei in the
neutron-rich side would act as a useful reference system to explore
the role of the Coulomb interaction on the exotic structure
formation. For this purpose, in this work we investigate the
structures of five pair of mirror nuclei $^{17}O$-$^{17}$F,
$^{26}$Na-$^{26}$P, $^{27}$Mg-$^{27}$P, $^{28}$Al-$^{28}$P and
$^{29}$Si-$^{29}$P which are all in the $2s_{1/2}$ state in the
framework of RMF. Combining with the RMF results of the 2$s_{1/2}$
states of $^{15}$O-$^{15}$N and $^{21}$Ne-$^{21}$Na \cite{Chen05},
we find that for the seven pairs of mirror nuclei, the exotic
structure of proton-rich nucleus is more obvious than that of its
mirror nucleus, the difference of exotic size between the mirror
nuclei becomes smaller with the increase of mass number A of the
nucleus, both the ratios of the valence proton and the valence
neutron root-mean-square (RMS) radius to the matter RMS radius in
each pair of mirror nuclei decrease linearly with the increase of
mass number A. In addition, in order to interpret the results above,
we calculate in detail two opposite effects of energy level shifted
and the Coulomb barrier hindrance on the exotic structure formation
by means of the single particle model (SPM) and find that the
contribution of the energy level shift is more important than that
of Coulomb barrier when A is small. However, the hindrance of
Coulomb barrier becomes more obvious with the increase of A. When A
is larger than 39, Coulomb effects on the exotic structure formation
will almost become zero because its two effects counteract with each
other.

\section{The RMF results and discussions}

As the relativistic mean-field (RMF) theory is a standard method for
describing properties of the spherical nuclei and some deformed
nuclei and its details can been found elsewhere such as Refs.
\cite{Wale74, Horo81, Sero86, Rein86, Rufa88, Rein89, Tani92,
Hira93, Lath94, Ring96, Lala97, Sero97, Meng98, Meng99, Paty99,
Bend03, Li04}, here we only describe the outline of the theory. For
a system with the interacting nucleons, $\sigma$, $\omega$, and
$\rho$ mesons and photons, the Lagrangian density is written as
\begin{eqnarray}
\mathcal {L} &=& \bar \psi [i\gamma ^\mu  \partial _\mu   - m -
g_\sigma  \sigma  - g_\omega  \gamma ^\mu  \omega _\mu   - g_\rho
\gamma ^\mu  \vec \tau  \cdot \vec \rho _\mu \nonumber\\
&& - e\gamma ^\mu \frac{{1 - \tau _3 }}{2}A_\mu  ]\psi\nonumber
+\frac{1}{2}\partial ^\mu \sigma \partial _\mu  \sigma  -
\frac{1}{2}m_\sigma ^2 \sigma ^2 \nonumber\\
&&- \frac{1}{3}g_2 \sigma ^3 - \frac{1}{4}g_3 \sigma ^4-
\frac{1}{4}\omega ^{\mu \nu } \omega _{\mu \nu }  +
\frac{1}{2}m_\omega ^2 \omega ^\mu \omega _\mu\nonumber\\
&&- \frac{1}{4}\vec \rho ^{\mu \nu }  \cdot \vec \rho _{\mu \nu } +
\frac{1}{2}m_\rho ^2 \vec \rho ^\mu   \cdot \vec \rho _\mu -
\frac{1}{4}A^{\mu \nu } A_{\mu \nu },
\end{eqnarray}
where
\begin{eqnarray}
\omega ^{\mu \nu } &=&\partial ^\mu  \omega ^\nu   - \partial
^\nu\omega ^\mu,\\
 A^{\mu \nu }  &=& \partial ^\mu  A^\nu   - \partial
^\nu  A^\mu,\\
\vec \rho ^{\mu \nu } & =& \partial ^\mu  \vec \rho ^\nu-
\partial ^\nu  \vec \rho ^\mu   - 2g_\rho  (\vec \rho ^\mu   \times
\vec \rho ^\nu),
\end{eqnarray}
The nucleon field and the rest mass are denoted as $\psi$ and $m$,
respectively. The meson fields and their masses are denoted by
$\sigma$, $\omega$, $\rho$ and $m_\sigma$, $m_\omega$, $m_\rho$,
respectively. The photon field $A_\mu$ produces the electromagnetic
interaction, and $e$ is its coupling constant. $g_\sigma$,
$g_\omega$, $g_\rho$ are the coupling constants between the mesons
and nucleons respectively, $g_2$ and $g_3$ are the non-linear
coupling constants of the $\sigma$ meson. $\tau_3$ is the third
component of the isospin Pauli matrices, i.e., $\tau _3 \left| n
\right\rangle  = \left| n \right\rangle$ and $ \tau _3 \left| p
\right\rangle  =  - \left| p \right\rangle$. Using Euler-Lagrang
equation, a set of coupled equations for nucleons, mesons and
photons can be obtained from the Lagrangian density function, which
can been solved self-consistently by iteration under the mean-field
approximation. After the final solutions are obtained, some
quantities we need, such as the binding energy, single-particle
levels, root-mean-square (RMS) radii of neutron and proton density
distributions, and so on, can be calculated from the wave functions.
In the effective lagrangian density function of the relativistic
mean-field $m$ is the mean value of the rest masses of proton and
neutron, $e$ satisfies $e^2/4\pi =1/137$ and $m_\sigma$, $m_\omega$,
$m_\rho$, $g_\sigma$, $g_\omega$, $g_\rho$,  $g_2$, $g_3$ are free
parameters. There are several well-tested nonlinear RMF parameter
sets NL1, NL2, NL3 and NL-SH which were obtained by fitting the
experimental observables, such as the binding energies and radii of
the nuclei \cite{Wale74, Horo81, Sero86, Rein86, Rufa88, Rein89,
Tani92, Hira93, Lath94, Ring96, Lala97, Sero97, Meng98, Meng99,
Paty99, Bend03, Li04}. For usual RMF calculation there are two
methods using spherical coordinate system and cylindrical coordinate
system, which are used in calculations of spherical nuclei and
reformed nuclei, respectively.

In this work five pairs of mirror nuclei $^{17}$F-$^{17}$O,
$^{26}$P-$^{26}$Na, $^{27}$P-$^{27}$Mg, $^{28}$P-$^{28}$Al and
$^{29}$P-$^{29}$Si and their core nuclei $^{16}$O, $^{25}$Na,
$^{25}$Si, $^{26}$Mg, $^{26}$Si, $^{27}$Al, $^{27}$Si, $^{28}$Si are
calculated by RMF. In the calculations the parameter set NL1
\cite{Rein86, Rein89} is chosen for $^{16}$O, $^{17}$F, and
$^{17}$O, and NL3 \cite{Lala97,Paty99} are chosen for other nuclei,
and the Pauli blocking effects are considered. In addition, all
nuclei are considered as spherical, which means the deformation
effect is omitted to simplify the calculations.

\begin{widetext}
\begin{center}
\begin{table}
\caption{The RMF results of five pairs of mirror nuclei
$^{17}$O-$^{17}$F, $^{26}$Na-$^{26}$P, $^{27}$Mg-$^{27}$P,
$^{28}$Al-$^{28}$P and $^{29}$Si-$^{29}$P and their core
nuclei.}\label{t1}
% Give a unique label
% For LaTeX tables use
\begin{tabular}{cccccccccc}
\hline \hline
  &$B_ {\mbox {exp.}}$ (MeV)& $B_{\mbox {the.}}$ (MeV)&$ R_n$(fm)&$R_p$(fm)&$R_m$(fm)&$R_{LN}$(fm)&
  $\varepsilon_{LN}$(MeV)&$R_{LN} /R_m$ &$\vert R_n-R_p \vert$(fm)\\
  \hline
  $^{16}$O&127.62&127.15&2.64&2.66&2.65&&&&0.02\\
  $^{17}$O&130.89&130.20&2.90&2.67&2.79&4.40&-3.49&1.58&0.23\\
  $^{17}$F&127.72&127.18&2.66&2.99&2.84&4.77&-0.33&1.68&0.33\\
  $^{25}$Na&202.53&199.03&2.94&2.80&2.88&&&&0.14\\
  $^{26}$Na&208.15&205.71&3.05&2.82&2.95&4.01&-7.04&1.36&0.23\\
  $^{25}$Si&187.01&183.88&2.80&3.03&2.93&&&&0.23\\
  $^{26}$P&187.15$\#$&185.45&2.82&3.13&3.00&4.29&-1.67&1.43&0.31\\
  $^{26}$Mg&216.68&211.71&2.92&2.86&2.89&&&&0.06\\
  $^{27}$Mg&223.12&219.29&3.02&2.87&2.96&3.94&-7.97&1.33&0.15\\
  $^{26}$Si&206.05&201.55&2.84&2.97&2.91&&&&0.13\\
  $^{27}$P&206.94&204.27&2.88&3.10&3.00&4.17&-2.53&1.39&0.22\\
  $^{27}$Al&224.95&219.78&2.90&2.88&2.89&&&&0.02\\
  $^{28}$Al&232.68&228.32&3.00&2.90&2.95&3.88&-8.81&1.32&0.10\\
  $^{27}$Si&219.36&214.12&2.85&2.94&2.90&&&&0.09\\
  $^{28}$P&221.42&217.19&2.87&3.05&2.97&4.08&-3.20&1.37&0.18\\
  $^{28}$Si&236.54&230.78&2.88&2.92&2.90&&&&0.04\\
  $^{29}$Si&245.01&240.25&2.98&2.93&2.96&3.83&-9.72&1.29&0.05\\
  $^{29}$P&239.29&234.68&2.90&3.03&2.97&4.00&-4.02&1.35&0.13\\
\hline\hline
\end{tabular}
% Or use
%\vspace*{5cm}  % with the correct table height
\end{table}
\end{center}
\end{widetext}

\begin{figure}[!h]%[tpb]
\begin{center}
\includegraphics[width=12cm,angle=0]{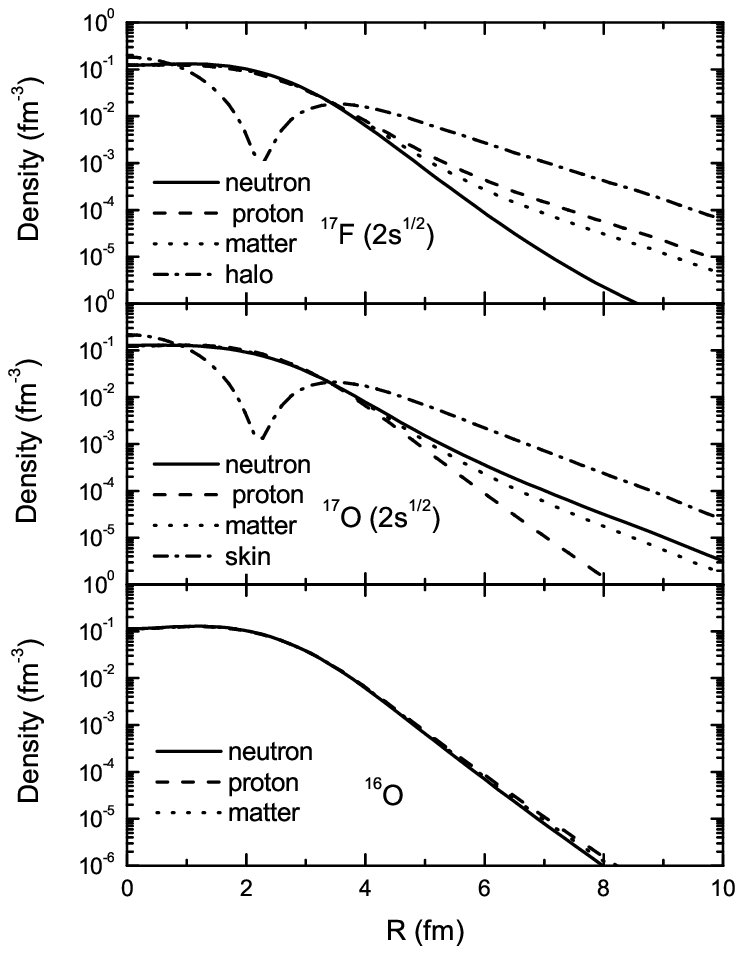} \label{fig1}
\caption{The density distributions of neutron, proton, matter, and
the last nucleon in $^{17}$O-$^{17}$F in the first excited state.
Solid, dashed, dotted, and dash-dotted curves are the density
distributions of protons, neutrons, matter and halo proton,
respectively. The density distributions of their core nucleus
$^{16}$O is also drawn for comparing.}
\end{center}
\end{figure}

The RMF calculation results for the 2$s_{1/2}$ states of five pairs
of mirror nuclei and their relative core nuclei are listed in Table
\ref{t1}. $B_ {\mbox {exp}}$ and $B_{\mbox {the}}$ are the
experimental and the calculated binding energies, respectively.
$R_n$, $R_p$, $R_m$ and $R_{LN}$ denote the calculated RMS radii of
neutron, proton, matter and the last valence nucleon density
distributions, respectively, $\varepsilon_{LN}$ is the
single-particle energy of the last nucleon, $R_{LN} /R_m$ is the
ratio of the valence nucleon RMS radius to the matter radius, and
$\vert R_n-R_p \vert$ represents the difference between the proton
and the neutron RMS radii. In Table  \ref{t1}, $B_{exp}$ are taken
from Ref.\cite{Audi93}, in which $B_{exp}$ of the nucleus $^{26}$P
is the estimated value (denoted as $\#$) by Audi and Wapstra because
its experimental binding energy is unknown. The density
distributions of neutron, proton, matter and the valence nucleus in
the mirror nuclei $^{17}$F-$^{17}$O which are calculated by the RMF,
are shown in Fig.1. Here the normalized density distributions $\rho
(r)$, which satisfies $\int_0^{ + \infty } {\rho (r)} r^2 dr = 1$,
is shown in order to compare the relative distributions of protons
and neutrons. Based on Table \ref{t1} and Fig. 1, the following
discussions are made and some important conclusions are obtained.

First, for the double magic nucleus $^{16}$O and its neighbor nuclei
$^{17}$O and $^{17}$F, the difference between the theoretical
binding energy $B_{the}$ and the corresponding experimental value
$B_{exp}$ is very small, and $B_{the}$ is at  most 0.5\% off. For
other four pairs of mirror nuclei and their core nuclei, there are
good agreement between their $B_{the}$ and $B_{exp}$, in which for
four pairs of mirror nuclei, $B_{the}$ is at most 2\% off, and for
other core nuclei $B_{the}$ is at  most 2.5\% off. These results
show that it is reasonable and reliable for us to use the spherical
RMF theory to describes the properties and structures of the nuclei
considered.

Second, it can be seen from Fig. 1 that there are long tails in the
density distribution of the valence proton of $^{17}$F and the
valence neutron of $^{17}$O, compared with their core nucleus. The
similar distributions also exist in other four pairs of mirror
nuclei. These show that all of the five pairs of the mirror nuclei
have exotic structures of halo or skin.

Third, there are obvious differences in exotic size of each pair of
mirror nuclei. For example, the binding energy $\varepsilon_{LN}$ of
the valence proton is much lower than that of the valence neutron in
its mirror nucleus, and the RMS radium $R_{LN}$ and $|R_p - R_n|$ of
the valence proton is larger than that of the valence neutron in its
mirror nucleus. In addition, it can also be seen from Fig.1 that the
tails in the density distribution of the valence proton of $^{17}$F
are longer than that of the valence neutrons of its mirror nuclei
$^{17}$O. The similar results occur in other four pairs of mirror
nuclei.  These show the exotic structure of valence proton is more
obvious than that of the valence neutron of its mirror nucleus.

Fourth, it can be seen from the change of the values $R_{LN}/R_m$ of
the five pairs of mirror nuclei in Table \ref{t1} that the
 values $R_{LN}/R_m$ of the valence protons and the valence neutrons
in each pair of the mirror nuclei all decrease with increase of mass
number A. Also, the difference of exotic size between each pair of
mirror nuclei becomes smaller with the increase of mass number A of
the nucleus. In order to compare the size of the exotic structure
between mirror nuclei more clearly, we plot $R_{LN} /R_m$ as a
function of the mass number A in Fig. 2. The RMF results in the
$2s_{1/2}$ state of the other two pairs of mirror nuclei
$^{15}$O-$^{15}$N and $^{21}$Ne-$^{21}$Na \cite{Chen05} are also
shown in this figure for comparison. It can be seen from Fig.2 that
the values of $^{15}$N, $^{17}$F, $^{21}$Na, $^{26}$P, $^{27}$P,
$^{28}$P and $^{29}$P are consistently larger than those of their
respective mirror partners, and the values $R_{LN} /R_m$ of the
valence-proton nuclei and the valence-neutron nuclei are almost in
two different lines and decrease gradually with the increase of mass
number A. The two linear functions obtained with the least square
fit are $(R_{LN} /R_m)_p = 2.16176 - 0.02824A$ for the valence
proton nuclei (valence proton line) and $(R_{LN} /R_m)_n = 1.95596 -
0.02302A$ for the valence neutron nuclei (valence neutron line),
respectively. The difference between the mirror nuclei decreases
with the increase of mass number $A$, so as two lines intersect at
about A=39. In addition, it is clear that the difference between the
valence proton line and the valence neutron line also decreases
linearly with increase of mass number A. This difference line (RMF
Coulomb line) corresponds to the contribution of the pure Coulomb
effect on the valence proton dispersion.

\begin{figure}[!h]%[tpb]
\begin{center}
\includegraphics[width=9.5cm,angle=0]{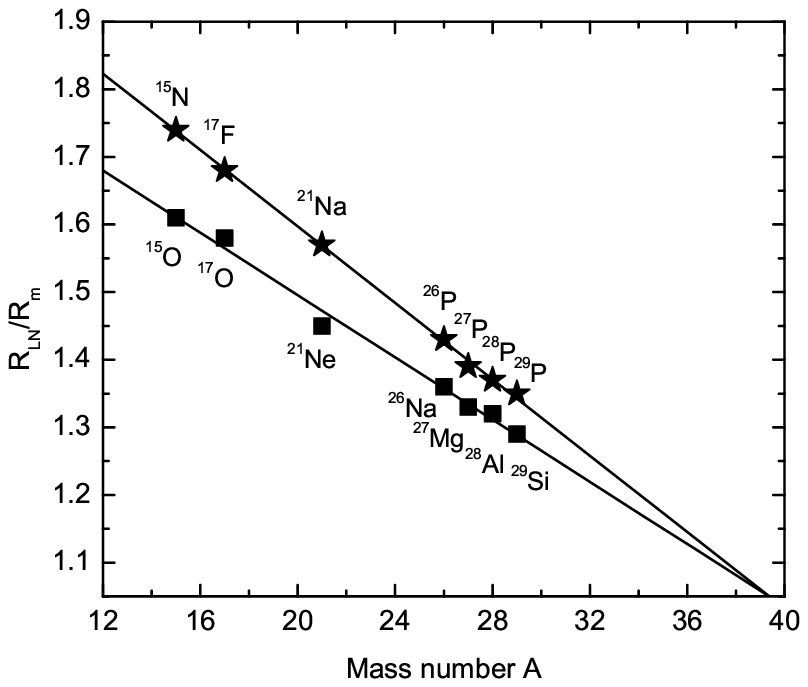} \label{fig2}
\caption{The ratios of the last nucleon RMS radius to the matter one
as a function of mass number in mirror nuclei. The solid stars stand
for the ratios in the proton-rich nuclei and the solid squares stand
for that in the neutron-rich nuclei. The results for
$^{15}$O-$^{15}$N and $^{21}$Ne-$^{21}$Na are taken from Ref.
\cite{Chen05}. The lines are the least square fit for the data.}
\end{center}
\end{figure}

\begin{figure}[!h]%[tpb]
\begin{center}
\includegraphics[width=10cm,angle=0]{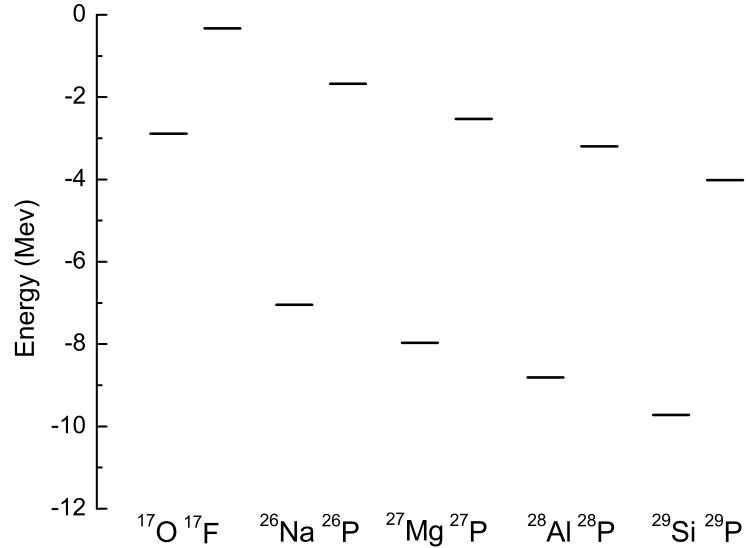} \label{fig3}
\caption{The single neutron energy levels of 2$s_{1/2}$ states for
$^{17}$O, $^{26}$Na, $^{27}$Mg, $^{28}$Al, $^{29}$Si and the single
proton energy levels 2$s_{1/2}$ states for $^{17}$F, $^{26}$P,
$^{27}$P, $^{28}$P, $^{29}$P.}
\end{center}
\end{figure}

\begin{figure}[!h]%[tpb]
\begin{center}
\includegraphics[width=12cm,angle=0]{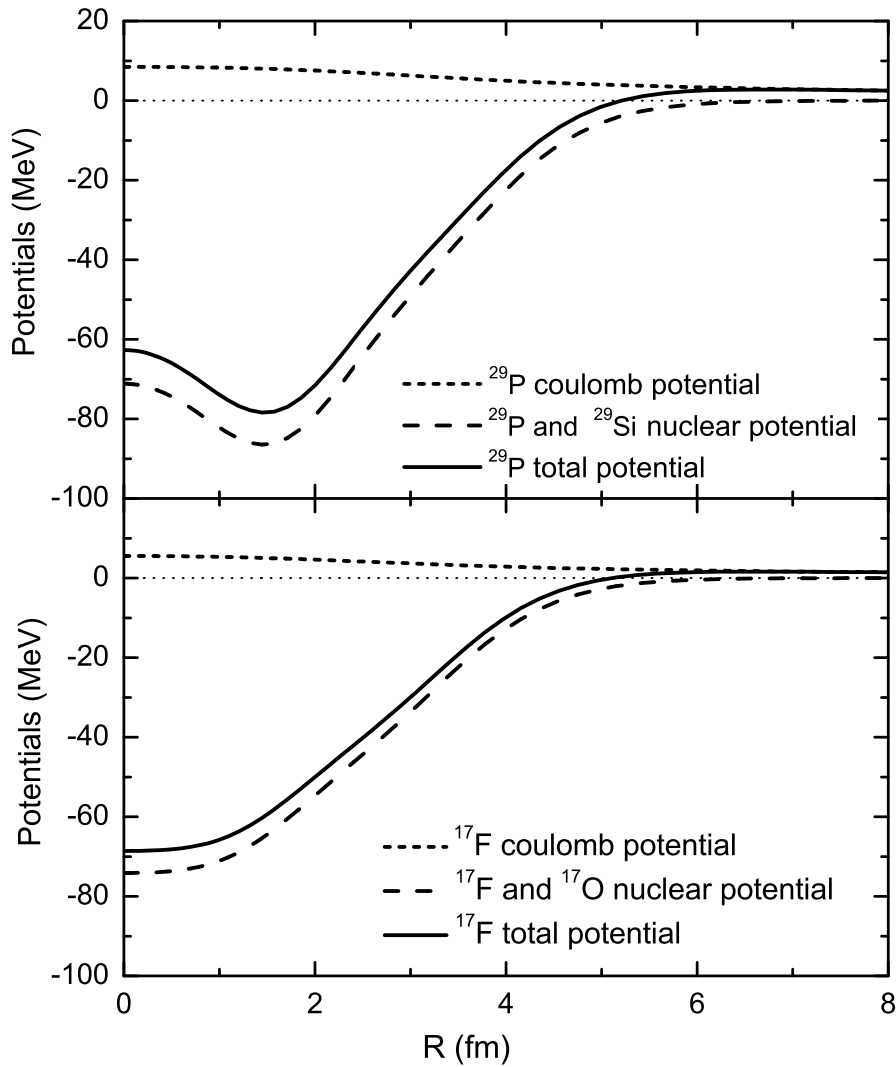} \label{fig4}
\caption{The variations of the means-field potentials with the
radial coordinates for $^{17}$O-$^{17}$F and $^{29}$Si-$^{29}$P.
Short-dashed, dashed and solid curves are the Coulomb potential,
nuclear potential and total potential.}
\end{center}
\end{figure}

\begin{figure}[!h]%[tpb]
\begin{center}
\includegraphics[width=12cm,angle=0]{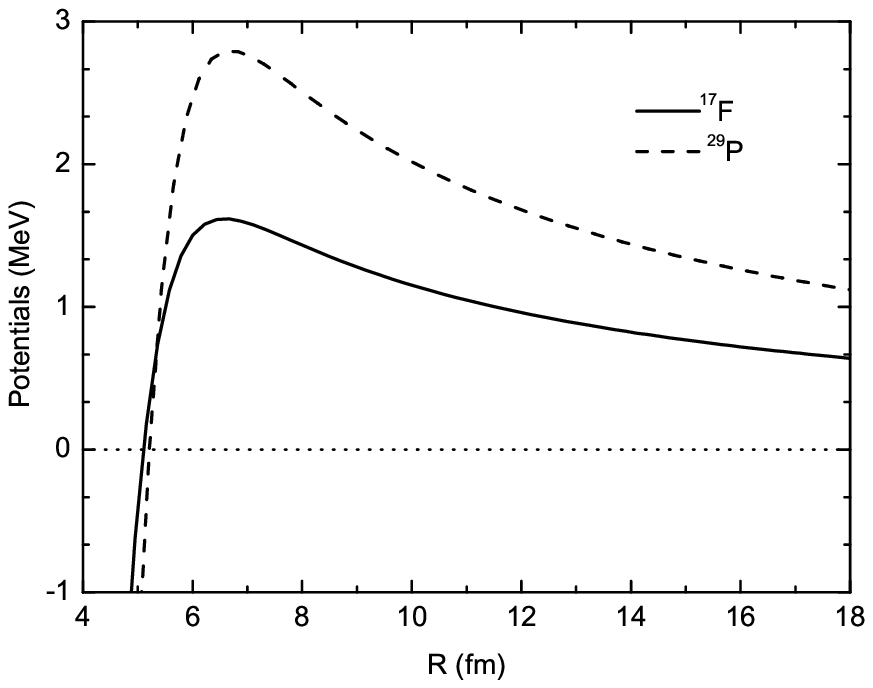} \label{fig5}
\caption{The Coulomb barriers of $^{17}$F and $^{29}$P. The solid
and dashed curves are for $^{17}$F and $^{29}$P, respectively.}
\end{center}
\end{figure}

The above results can be explained qualitatively  by the RMF
calculation results. Because the nuclear force is approximately
charge-independent and there are the same nuclear force interaction
in each pair of mirror nuclei, the difference between the structures
of mirror nuclei should only come from the Coulomb interaction. The
Coulomb effects on valence proton distribution include two opposite
actions. One action is that it makes the energy level shift closer
to the Fermi level, which means the valence proton is easier to
extend far from the nuclear potential. The other action is that the
Coulomb barrier hinders the diffusion of the valence proton and
formation of exotic structure. We plot the single nucleon energy
levels of 2$s_{1/2}$ states of five pairs of the mirror nuclei in
Fig. 3 and the nuclear, the Coulomb and the total potentials for
$^{17}$O-$^{17}$F and $^{29}$Si-$^{29}$P in Fig. 4. The potentials
for the other three pairs of mirror nuclei are similar to that of
$^{17}$O-$^{17}$F and $^{29}$Si-$^{29}$P. It can be seen from Fig.3
that the single proton energy levels in $^{17}$F, $^{21}$Na,
$^{26}$P, $^{27}$P, $^{28}$P and $^{29}$P are higher than the
corresponding single neutron energy levels in their mirror nucleus
$^{17}$O, $^{26}$Na, $^{27}$Mg, $^{28}$Al and $^{29}$Si, which is
induced by the Coulomb interaction. It can also be seen from Fig. 3
that the energy level shifts in the 2$s_{1/2}$ states between mirror
nuclei become larger with the increase of mass number A except
$^{17}$O and $^{17}$F. On the other hand, the calculation results
shown that the hights of Coulomb barrier increase with the increase
of mass number A.  The Coulomb barriers of $^{17}$F and $^{29}$P are
shown in Fig.5. The Coulomb barrier of $^{26}$P, $^{27}$P, and
$^{28}$P lie between those of $^{17}$F and $^{29}$P (not shown in
the figure). The two opposite effects of Coulomb interaction
determine the contribution of Coulomb potential to the valence
proton dispersion together. Although the RMF theory cannot separate
quantitatively the two opposite effects of Coulomb interaction to
valence proton, it can be inferred from Fig.4 and 5 that when mass
number A is small, the contribution of the energy level shift is
more important than that of the Coulomb barrier, but the hindrance
of the Coulomb barrier becomes more obvious with the increase of A.
When A is about 39, Coulomb effects on the exotic structure
formation will almost become zero because its two effects counteract
with each other. This qualitative conclusion will be proven
quantitatively in the following single-particle model (SPM)
analysis.

\section{The SPM results and discussions}
\begin{widetext}
\begin{center}
\begin{table}
\caption{The contributions of the energy level shifted and the
Coulomb barrier to exotic formation calculated with the
single-particle model.}
\label{tab2}       % Give a unique label
% For LaTeX tables use
\begin{tabular}{ccccccccccccc}
\hline \hline
  &$^{17}$O&$^{17}$F&$^{21}$Ne&$^{21}$Na&$^{26}$Na&$^{26}$P&$^{27}$Mg&$^{27}$P&
$^{28}$Al&$^{28}$P&$^{29}$Si&$^{29}$P\\
  $R_{LN}$ (fm)&4.27&5.43&4.40&5.37&4.19&4.87&4.10&4.60&3.98&4.33&3.93&4.22\\
  $R_{i}$ (fm)&&6.68&&9.49&&5.10&&2.23&&1.24&&0.98\\
  $R_{d}$ (fm)&&-5.52&&-8.52&&-4.42&&-1.73&&-0.89&&-0.39\\
  $R_c$ (fm)&&1.16&&0.97&&0.68&&0.50&&0.35&&0.29\\
 \hline\hline
\end{tabular}
% Or use
%\vspace*{5cm}  % with the correct table height
\end{table}
\end{center}
\end{widetext}

\begin{figure}[!h]%[tpb]
\begin{center}
\includegraphics[width=12cm,angle=0]{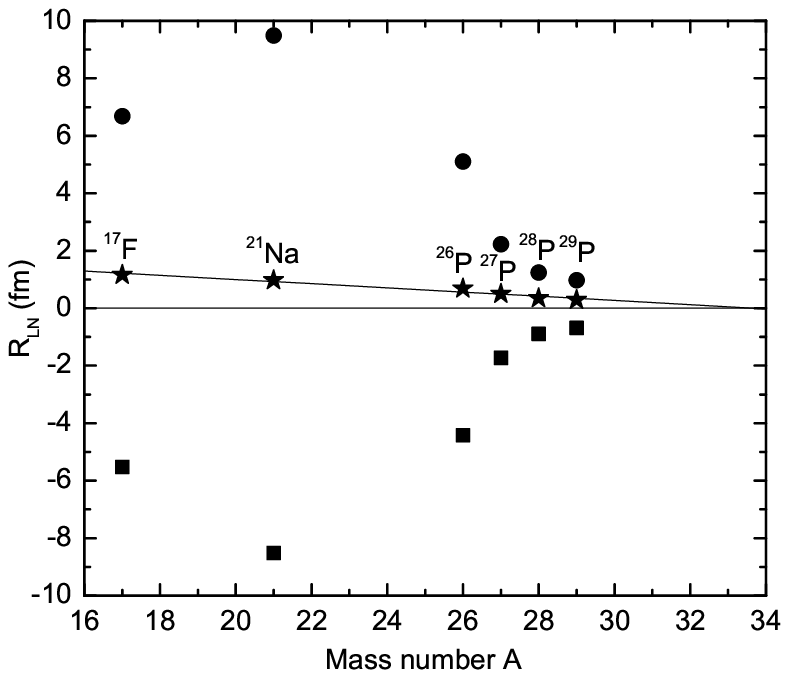} \label{fig6}
\caption{The contributions of the energy level shift and the Coulomb
barrier to exotic formation calculated with the single-particle
approach. The solid circles  denote the increase of $R_{LN}$ induced
by the energy level shifted, the solid squares denote the decrease
of $R_{LN}$ induced by the Coulomb barrier and the solid stars
denote the total Coulomb effect on $R_{LN}$.}
\end{center}
\end{figure}

In the single-particle model, the nucleus is assumed to be composed
of the nuclear core and the valence nucleon outside the core. The
normalized single-particle radial wave function in the ($n l j$)
bound state $\phi _{nlj} (r)$ can be obtained by solving the
Schr\"{o}dinger equation. The potential is chosen as
\begin{eqnarray}
V(r) = V_N (r) + V_C (r) ,
\end{eqnarray}
where the nuclear potential $V_N (r)$  is chosen as Woods-Saxon
potential
\begin{eqnarray}
V_N (r) = V_0 /\left\{ {1 + \exp (\frac{{r - r_0 A^{1/3} }}{{a_0
}})} \right\},
\end{eqnarray}
and $V_C (r)$  is Coulomb potential. Here the potential depth $V_0$
of Woods-Saxon potential is adjusted to reproduce the valence
nucleon separation energy, $r_0$ and $a_0$ are the radius and
diffuseness parameter, respectively. In this work, $r_0$ and $a_0$
are chosen as the normal values 1.25 fm and 0.65 fm, respectively.
The root-mean-square (rms) radius of the distribution of the last
nucleon can be obtained from the single-particle radial wave
function $\phi _{nlj} (r)$ by
\begin{eqnarray}
R_{LN}= \left[ {\int_0^\infty {r^4 \phi _{nlj}^2 (r)dr} }
\right]^{1/2}.\label{eq8}
\end{eqnarray}

The numerical results with the SPM are listed in Table \ref{tab2},
where $R_{i}$ denotes the increase of $R_{LN}$ induced by the energy
level shifted and $R_{d}$ denotes the decrease of $R_{LN}$ induced
by the Coulomb barrier. $R_c$ denotes the contribution of total
Coulomb effect  on the valence proton dispersion and $R_c = R_i +
R_d$. We select $^{17}$O- $^{17}$F as an example to explain the
details of the calculations. At first, we obtain the RMS radii of
the valence nucleon with the single-particle model (4.27 fm for
$^{17}$O and 5.43 fm for $^{17}$F). Next, we assume that the binding
energy of $^{17}$O is the same as $^{17}$F and calculate the valence
neutron RMS radii for $^{17}$O under this condition. The result is
10.95 fm, which is much larger than the actual value of $^{17}$O.
This implies that the energy level shifted makes the RMS radius
increase 6.68 fm if the effects of Coulomb barrier switch off.
However, the Coulomb barrier hinders the exotic formation in
reality. The RMS radius of $^{17}$F is less than the result obtained
by neglecting the effect of the Coulomb barrier. Therefore, the
Coulomb barrier makes RMS radii decrease 5.52 fm. In Table
\ref{tab2}, "-" of $R_d$ values shows decrease of the valence proton
dispersion. It shows that two effects of the Coulomb interaction are
opposite on the formation of exotic structures of valence proton,
and the contributions of the energy level shifted are more important
than that of the Coulomb barrier. The same calculations are applied
to the other pairs of mirror nuclei. In order to show these rules
more clearly, the variations of $R_{i}$, $R_{d}$ and $R_{c}$ with
the mass numbers are given in Fig.6. It is seen from Table II and
Fig.6 that the two effects of the Coulomb interaction on the valence
proton dispersion are opposite, and when mass number A is smaller
the contribution of the energy level shifted is more important than
that of the Coulomb barrier hindrance. However, the hindrance of the
Coulomb barrier becomes more obvious with the increase of A, and
total effect of the Coulomb interaction on the valence proton
dispersion decreases linearly with increase of the mass number A of
the nucleus, and when A is larger than about 34, Coulomb effects on
the exotic structure formation will almost become zero because its
two effects counteract with each other. These conclusions are in
good agreement with those obtained with RMF although there are some
differences in their values, which maybe come from the difference
between the two theoretical models RMF and SPM.

\section{Summary}

In summary, we have investigated the exotic structures in $2s_{1/2}$
state of five pairs of mirror nuclei $^{17}$O-$^{17}$F,
$^{26}$Na-$^{26}$P, $^{27}$Mg-$^{27}$P, $^{28}$Al-$^{28}$P and
$^{29}$Si-$^{29}$P with RMF and SPM in order to explore the role of
the Coulomb interaction on the proton halo formation. By analyzing
the RMF results, we find that the exotic structure of valence proton
is more obvious than that of the valence neutron of its mirror
nucleus, the $R_{LN} /R_m$ values of the valence proton and the
valence neutron of each pair of mirror nuclei decrease linearly with
the increase of mass number A of the mirror nuclei, and the
difference between the values of each pair of mirror nuclei becomes
smaller linearly with the increase of mass number A. When A is about
39, the difference closes zero. By analyzing quantitatively two
opposite effects of the Coulomb interaction on the valence proton
dispersion in  some mirror nuclei with SPM, we find that the
contributions of the energy level shift are more important than that
of the Coulomb barrier, and the total Coulomb effect on the valence
proton dispersion becomes smaller linearly with increase of the mass
number A, and it closes zero when A is larger than 34, which means
two effects of Coulomb interaction on valence proton counteract with
each other. These results basically agree with the conclusions
obtained by RMF.

\section*{ACKNOWLEDGMENTS}

This work is supported by the National Natural Science Foundation of
China under Grant Nos 10604008 and 10435020, A foundation for the
Author of National Excellent Doctoral Dissertation of China, and
Beijing Education Committee Under Grant No. XK100270454.

\end{document}